\documentclass[aps,prb,a4paper,twocolumn,showpacs,floatfix,citeautoscript]{revtex4}
\usepackage{graphicx}
\usepackage{amsfonts}
\usepackage{amsmath}
\usepackage{amssymb}
\usepackage{bm}
\usepackage{dcolumn}
\usepackage{hyperref}
                     \usepackage[usenames]{color}

\hypersetup{pdfborder=0 0 0,colorlinks=true,citecolor=blue,linkcolor=blue}
\begin{document}

\title{Double quantum dot as a minimal thermoelectric generator}
\author{S. Donsa$^1$, S. Andergassen$^2$, and K. Held$^1$}
\affiliation{$^1$ Institute of Solid State Physics, Vienna University of Technology, A-1040 Vienna, Austria}
\affiliation{$^2$ Faculty of Physics, University of Vienna, Boltzmanngasse 5,  A-1090 Vienna, Austria}

\date{\today}

\begin{abstract} 
Based on numerical renormalization group calculations,
we demonstrate that  experimentally realized double quantum dots
 constitute a minimal thermoelectric generator.
In the Kondo regime, 
one quantum dot acts as an n-type and the other one as a p-type thermoelectric device. Properly connected a capacitively coupled double quantum dot provides a miniature power supply utilizing the thermal energy of the environment.
\end{abstract}

\maketitle
\section{Introduction}
The theoretical prediction\cite{Glazman1} and subsequent experimental
confirmation\cite{Goldhaber-Gordon1} of the Kondo effect in a
%single 
quantum dot represents, without doubt, a 
prominent research highlight in the field of nanoscopic physics.
While the Kondo effect of impurities in bulk metals induces a reduction of the conductance\cite{deHaas36}, %the (additional) 
its nanoscopic realization leads to an enhancement due to quantum many-body effects.
%Kondo resonance at the Fermi level makes an otherwise Coulomb blockaded quantum dot conducting.
More recently, the Kondo effect was also observed in double quantum dots\cite{Borda1,Sasaki1,Huebel2,ddot}. Here, the two quantum dots may
act as a pseudospin in analogy to  spin-up and -down
for the usual Kondo effect in a single
quantum dot. Combined, spin and pseudospin can give rise
to an enhanced SU(4) Kondo effect. Such double quantum dots are particularly
exciting from the fundamental research point of view 
since (pseudo-)spin-up and -down can be controlled separately and their conductance can be measured independently\cite{Huebel1}.

Besides for their electrical conductance,  quantum dots are also considered as potential solid state energy converters\cite{mahan,1dotnoU,correltherm,finiteV}. 
The high degree of tunability of nanoscale devices allows them to be operated at
optimal thermoelectric efficiency. Promising in this respect are  
ultrasharp resonances which can be achieved through the Kondo effect\cite{thermodots,key-1,CK,hdots,correltherm1}. 
However, for the single-dot Kondo effect, the
resonance is centered around the Fermi level so that electron and hole
contributions cancel. The thermopower is vanishingly small\cite{key-1}.
A possibility to move the resonance away from the Fermi level is  applying an 
external magnetic field. This  splits the Kondo
resonance with one spin-species above and the other below the Fermi energy, but
the total (spin-averaged) thermopower  stays small\cite{CK,hdots}. Another idea 
has been to employ the charge Kondo effect, realized in an Anderson impurity model with attractive interaction\cite{CK}. This is however difficult
to realize experimentally. It requires e.g.\ a strong electron-phonon coupling to realize an effective  interaction that is attractive.

\begin{figure}[t!]
\includegraphics[width=6cm,angle=270]{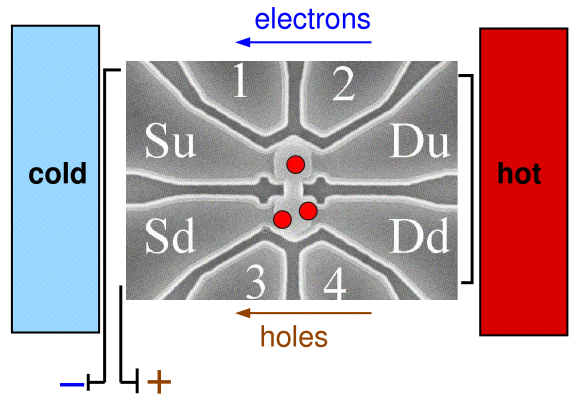}
\caption{(Color online) Schematic figure of %how to make a thermoelectric device out of  a double quantum dot. 
the considered thermoelectric double quantum dot device.
For suitable gate voltages of the electrodes 1,2 and 3,4, a negative current of (quasi-)electrons flows in the up quantum dot and a positive current (quasi-)holes in the down quantum dot. Hence, if the two dots are connected on the right hand side, the up and down quantum dot provide %serve as
on the left hand side the negative and positive pole of a power supply energized by excess heat from the environment
(the micrograph of the quantum dot at the center has been reproduced from Ref.~\onlinecite{Huebel1}.}
\label{Fig:Scheme}
\end{figure}

%In this letter, %!!! we may specify this once the paper is accepted
In this paper, we show that a  capacitively coupled double quantum dot in the Kondo regime ultimately overcomes these difficulties. Such double dots, which
are already realized experimentally,
 represent a stand-alone  thermoelectric generator, see Fig.\ \ref{Fig:Scheme}.
Given an external heat gradient, (quasi-)electrons and holes 
dominate the linear transport in the two respective quantum dots such that
a total current is generated if the quantum dots are suitably
connected. We stress that the same theoretical concept can be realized 
 in experimentally very distinct systems, such as  molecular transport\cite{molecularelectronics} and cold atoms\cite{coldatoms}.

\section{Model and method} Basis of our calculations is the experimentally realized  
double quantum dot of  Ref.\ \onlinecite{Huebel1}. The experimental signatures of the Kondo physics in the conductance  
are well described 
%was well reproduced 
by an Anderson impurity model\cite{Huebel2}
with a single spin-degenerate level in each dot:
\begin{eqnarray}\label{Hamiltonian}
    \hat{H}=&&\!\sum\limits_{i\in\left\{{\rm u,d}\right\}}\!\!\!\big(\varepsilon_i  \hat{n}_i+ U_i \hat{n}_{i\uparrow}\hat{n}_{i\downarrow}\big)
    + U' \hat{n}_{\rm u}\hat{n}_{\rm d}\nonumber\\
    &&+\!\!\sum\limits_{Ri,k,\sigma}\!\!\!\varepsilon_k 
    \hat{c}^{\dagger}_{Ri,k,\sigma}\hat{c}^{}_{Ri,k,\sigma}\nonumber\\
    &&+\!\!\sum\limits_{Ri,k,\sigma}\!\!\!\left(V_{Ri}  \hat{a}^{\dagger}_{i,\sigma}\hat{c}^{}_{Ri,k,\sigma}+ {\rm
    h.c.}\right)\;
\label{Eq:AIM}
\end{eqnarray}
Here, $\hat{n}_{i,\sigma}=\hat{a}^{\dagger}_{i,\sigma}\hat{a}^{}_{i,\sigma}$, where $\hat{a}^{\dagger}_{i,\sigma}$  ($\hat{a}_{i,\sigma}$) creates
(annihilates) an electron with spin
$\sigma\in\{\downarrow,\uparrow\}$ in dot $i\in\{{\rm u}, {\rm d}\}$ with energy level $\varepsilon_{\rm i}$. $U_i$ 
is the Coulomb interaction %between two electrons 
on quantum dot $i$ (for simplicity we use $U_i=U$ in the following), and $U'$ is the capacitive coupling between the two dots.
{Note that there is no direct tunneling between the two dots.} 
 $\hat{c}^{\dagger}_{Ri,k,\sigma}$ is the
creation operator for an electron in the source or drain lead $R\in\left\{{\rm
S,D}\right\}$   with
wavenumber $k$ and energy $\varepsilon_k$. The leads are
coupled to quantum dot $i$ by the hybridization $V_{Ri}$,
which corresponds to a  tunneling rate $\Gamma_{R i}=2\pi\rho |V_{R i}|^2$ for a constant lead density of states $\rho$ (which can be taken as lead independent since only the combination $\rho |V_{R i}|^2$ is of relevance). 

As in Ref.\ \onlinecite{Huebel2}, the numerical renormalization group (NRG)\cite{NRG} approach is used to calculate the spectral function $A_i(\omega)$ of the two quantum dots %.From $A_i(\omega)$ in turn, 
from which
the conductance $G_i$, the thermopower $S_i$, and the electronic contribution to the thermal
conductance $K^e_{i}$ %through quantum dot $i$ 
can be determined in linear response via the Meir-Wingreen formula\cite{Meir},
see Ref.\ \onlinecite{key-1}:
\noindent 
\begin{eqnarray}
G_i(T)&=&e^{2}I_i^{0}(T)\\
S_i(T) &=&-\frac{1}{|e|T}\frac{I^{1}_i(T)}{I^{0}_i(T)} \label{Eq:I}\\ 
K^e_i(T) &=&\frac{1}{T}\left[I^{2}_i(T)- \frac{I^{1}_i(T)^2}{I^{0}_i(T)} \label{Eq:K}\right]\,.
\end{eqnarray}
Here, $e$ is the elementary charge, $T$ the temperature, and $I^{n}_i$ denote the transport integrals
\begin{equation}
I^{n}_i(T)= \frac{2}{h}\int d\omega\:\omega^{n}{\cal T}_i(\omega) \left(-\frac{\partial f(\omega)}{\partial\omega}\right) \, ,\\
\end{equation}
where the transmission is given by
\begin{equation}
{\cal T}_i(\omega)=2\pi\frac{\Gamma_{L i}\Gamma_{R i}}{\Gamma_{L i}+\Gamma_{R i}}A_i(\omega)\, .
\end{equation}
The $I^{n}_i$ are thus the moments of the spectral function weighted by the derivative of the Fermi function $f(\omega)$
around the Fermi energy $\omega=0$. 
%previous:
%Note that, since the two
%quantum dots only interact through the interaction $U'$, there is no direct
%tunneling between them, see Eq.~(\ref{Eq:AIM}).

%new paragraph:
\section{General discussion}
For a better understanding, we consider the Sommerfeld expansion of the 
thermopower at low $T$:
\begin{equation}
\label{sf}
S_i(T)=-\frac{\pi^{2}k_{B}}{3|e|}k_{B}T\frac{d A_i(\omega)/d\omega|_{\omega=0}}{A_i(0)}\,.
\end{equation}
In the absence of a magnetic field $A_{i\uparrow}(\omega) = A_{i\downarrow}(\omega)  = A_{i}(\omega) $ is spin independent. 
A large thermopower 
is thus obtained for a highly asymmetric $A_i(\omega)$ with a large slope in the derivative
at the Fermi energy $d A_i(\omega)/d\omega|_{\omega=0}$ and a small $A_i(0)$. 
As we will see below, in the Kondo regime of a double quantum dot,
such an asymmetry with opposite slope for the two dots is indeed possible --- in contrast to the  single quantum dot case.

The physical constraints  for a large thermopower can be further elucidated by relating  $A_i(\omega)$ to the occupation $n_{i \sigma}$ of dot $i$ for spin $\sigma$ via the Friedel sum rule. The dot- and spin-resolved thermopower can hence be expressed as\cite{hewson}
\begin{equation}
S_{i \sigma}(T) =-\frac{\pi\gamma T}{|e|}\cot(\pi n_{i \sigma})
\end{equation}
with the linear coefficient of the specific heat $\gamma$. 
For a single quantum dot in the Kondo regime,  $n_{i \sigma}\sim 1/2$
so that $S_{i \sigma}\sim 0$ is vanishingly small\cite{key-1}. 
Applying an external magnetic field leads to,  say, $n_{i \uparrow}\!>\!1/2$ and
 $n_{i \downarrow}\!<\!1/2$. Hence,  $S_{i\uparrow}>  0$  and $S_{i\downarrow}<  0$,
but the total thermopower   $\sum_{\sigma} S_{i\sigma}$ remains small\cite{CK,hdots}. 
For the double quantum dot, we have a similar situation
but with the two distinct  dots ($i=u$ and $d$) now playing the role of the 
spin ($\uparrow$ and  $\downarrow$).
If there was not an additional spin-degree of freedom for the double dot, we would even have the same situation, except for one important difference: the two contributions
$i=u$ and $d$ are now spatially separated. Hence for a proper geometry,
   $S_{u\sigma}>  0$  and $S_{d\sigma}<  0$ can be even employed as 
the $p$- and $n$-type part of a thermoelectric device, see Fig.\  \ref{Fig:Scheme}.
We note that the  case of an {\it attractive} interaction\cite{CK} 
is related to the magnetic field situation, by a particle-hole
transformation for one spin species. For the particle-hole transformed spin species, this also changes the sign of
$S_{i \sigma}$; there is no cancellation in
 $\sum_{\sigma} S_{i\sigma}$.

\section{Results} 
\begin{figure}[t!]
\includegraphics[width=6cm,angle=270]{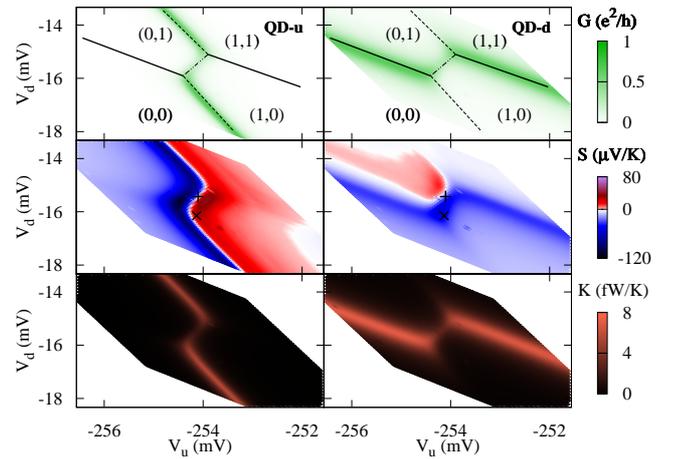}
\caption{(Color online)
Conductance $G$ %$G(V_{\rm u}, V_{\rm d})$ 
(upper panel), Seebeck coefficient $S$ (central panel),
and thermal conductivity $K^e$ (lower panel) for the double quantum dot of Ref.~\onlinecite{Huebel1} as a function of the applied gate voltages $V_{\rm u}$ and $V_{\rm d}$. The left (right) panels show the transport through the up (down) quantum dot as calculated by NRG, see text for the parameters.
%for the Anderson impurity model with two dots (impurities) and parameters given in the text.
The Kondo resonance develops in proximity of the
degeneracy line between  ($1$,$0$) and ($0$,$1$) electrons in the (up,down) quantum dot, see charge stability diagram in the upper panels. 
%If we move slightly off the 
Close to the 
maximum in the conductance, %e.g.\ for the dot in the central panel, 
the Seebeck coefficient of the up (down)
quantum dot is positive (negative) indicating charge transport of opposite
sign through the two dots.
\label{Fig2}
}
\end{figure}

We now analyze the thermoelectric properties, 
the conductance, the thermopower or Seebeck coefficient, and the thermal conductivity, 
of the double quantum dot of Ref.~\onlinecite{Huebel1}.
Figure \ref{Fig2} summarizes the NRG results obtained for the Anderson model described by \eqref{Eq:AIM}, with   $U' = 163\,\mu\,$eV, $U=620\,\mu\,$eV, $\Gamma_{{\rm Su}}=24\mu\,$eV, $\Gamma_{{\rm Du}}=8\mu\,$eV, $\Gamma_{{\rm Sd}}=38\mu\,$eV,  and $\Gamma_{{\rm Dd}}=21\mu\,$eV at $T=25\,$mK, see \onlinecite{Huebel2}. Here, the one-particle
energy levels $\varepsilon_i$ are related to the gate voltages applied to the up and down quantum dot by
\begin{eqnarray}\label{transformation}
\left(\!\!\begin{array}{c}{V_u}\\
{V_d}\end{array}\!\!\right)\!/{\rm mV} =
\left(\!\!\begin{array}{rr} -1.62& 1.74 \\
1.26&-3.19 \end{array}\!\!\right)\left(\!\!\begin{array}{c} \varepsilon_{\rm u}/U\\
\varepsilon_{\rm d}/U\end{array}\!\!\right)
+\left(\!\!\begin{array}{c}{-254.2}\\
{-15.6}\end{array}\!\!\right)\,.
\end{eqnarray}
The results for the conductance through both quantum dots (upper panel) reproduce %capture 
the experimental data as illustrated in Ref.~\onlinecite{Huebel1}.  Increasing the gate voltage $V_{\rm u}$ ($V_{\rm d}$) of the two electrodes  1,2 (3,4) in Fig. \ref{Fig:Scheme}, the number of electrons in the up (down) dot is increased by one; at the degeneracy point the Coulomb blockade 
is lifted leading to an enhanced conductance.
Due to the capacitive coupling of the dots, the gate voltages  $V_{\rm u}$ and $V_{\rm d}$  do not only affect the up and down quantum dot individually, but both dots. As a consequence, instead of a square-like pattern a characteristic honeycomb structure emerges in the charge stability diagram as depicted in the upper panel of Fig.\ \ref{Fig2}.

Since the Anderson impurity model describes the measured conductance very well,
we are confident that the theoretical results for the %other calculated 
thermoelectric properties %transport coefficients 
correspond to the actual experimental situation.
The central panels of Fig.  \ref{Fig2}  show the thermopower $S_i$ of the 
two quantum dots (left and right panel, respectively).
Negative (blue) values indicate the flow of negatively charged
(quasi-)electrons from the hot to the cold side of an external heat gradient, %bath
and positive (red) values the flow of positively charged
(quasi-)holes. At the maximal conductance (see upper panel),
the thermopower is rather small. Shifting the gate voltages
slightly away from this maximum, e.g., to the cross $\times$ indicated in Fig. \ref{Fig2}, a large thermopower develops whose values are particularly enhanced in the Kondo regime, for both quantum dots but with opposite sign.

Hence, as indicated in Fig.\ \ref{Fig:Scheme}, a thermal gradient induces
a current in opposite direction through the up and down quantum dot.
This means that the double quantum dot %already 
constitutes a thermoelectric device. Connecting the drain electrodes of the two quantum dots, i.e.,
employing an even simpler design were the two drain electrodes are substituted 
by a bar or large quantum dot, the two source electrodes ${\rm Su}$ and ${\rm Sd}$ in Fig.\ \ref{Fig:Scheme} act as the plus and minus source of a power supply.

\begin{figure}[t!]
\includegraphics[width=6.5cm,angle=270]{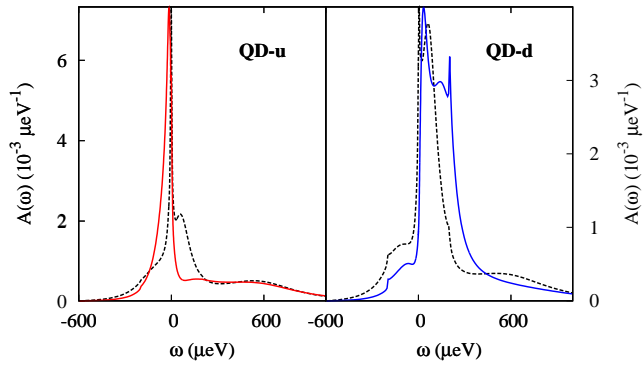}
\caption{(Color online)
Left (right) panel: 
Spectral function $A_{{\rm u}({\rm d})}$ of the up (down) quantum dot 
for $V_{\rm u}=-254.6\,$mV and $V_{\rm d}=-14.8\,$mV, corresponding to %the point
a pronounced thermopower ($\times$) in \protect  Fig.\ \ref{Fig2} (solid lines). The main spectral weight of the Kondo resonance is
below (above) the Fermi level at $\omega=0$. For comparison the spectral function for %in presence of  
a reduced thermopower ($+$) are shown (dashed lines).
\label{Fig3}}
\end{figure}

To further elucidate the understanding of the large thermopower of opposite sign, we show in  Fig.\ \ref{Fig3} the spectral functions of both quantum dots for the gate voltages indicated  %the dot 
in Fig.\ \ref{Fig2}.  We observe that the
sharp Kondo resonance of the two quantum dots is located directly  below and
above  the Fermi level for the up and down quantum dot, 
respectively. Hence, the spectral functions $A_i(\omega)$ are highly asymmetric around the Fermi level.
On the contrary, for a symmetric behavior around the Fermi level (dashed line in Fig.\ \ref{Fig3}) electrons and holes alike migrate from the hot to the cold side and their net current cancels,  $I_i^1\approx 0$.
In general we have such  a symmetric situation on the line separating the $(0,1)$ and $(1,0)$ occupation regions (see upper panel of Fig.\ \ref{Fig2}).
Changing the gate voltages perpendicular to this line leads to a splitting of the  $u$- and $d$-dot Kondo resonance, similar to applying a magnetic field for a single quantum dot. Note that due to its larger hybridization or tunneling rate $\Gamma_{Ld}+\Gamma_{Rd}$, the down dot has a much wider resonance and the situation is not completely symmetric.
%submitted version:
%This can be quantified through the Sommerfeld expansion of the 
%thermopower at low $T$ for frequency independent $\Gamma_{Ri}$
%\begin{equation}
%S_i(T)=-\frac{\pi^{2}k_{B}}{3|e|}k_{B}T\frac{d A_i(\omega)/d\omega|_{\omega=0}}{A_i(0)}\,.
%\end{equation}
%Here, the  Kondo effect leads to a sharp peak in  $A_i(\omega)$. This peak has to be off the Fermi energy ($\omega=0$) for a large $d A_i(\omega)/d\omega|_{\omega=0}$ and a small $A_i(0)$ in order to yield an enhanced thermopower. 
%For a single quantum dot with repulsive interaction, 
%a correlation-induced sharp peaks can occur away from
%half filling only for huge gate voltages of the order of the Coulomb interaction (due to the pinning of the energy levels to the Fermi energy), or in a magnetic field
%\cite{CK,hdots}. 
%At the order of the Kondo temperature $T_K$, a significant enhancement of the thermopower
%can be obtained only for {\it attractive} interactions, leading to the charge Kondo effect \cite{CK}. 
%In contrast, in a double dot 
%the dot occupancy can be easily manipulated
%by tuning the applied gate voltages individually, 
%providing a new route to the realization of a large thermopower in quantum dot systems.
 As a function of temperature, the maximum in the thermopower occurs on a temperature scale which correlates with the gate voltage and is therefore highly tunable. 
However, a systematic study of the temperature dependence has not been performed for the device considered here.

%As discussed above, the thermopower varies with the temperature. 
%However, a systematic study of the temperature dependence has not been performed for the device considered here.

An efficient thermoelectric element is characterized also by
%Another prerequisite for a good thermoelectric element  is 
a low thermal conductivity $K^e$ which determines the thermoelectric figure of merit defined % for quantum dots as
by\cite{key-1} $ZT= S^2T G/K$.	
Fig.\ \ref{Fig2} shows the electronic contribution to $K^e$
determined by Eq.~(\ref{Eq:K}). 
The double dot exhibits a maximum in the thermal conductivity 
%at symmetric gate voltages ???
in correspondence of a maximum in the conductance. A large thermopower is however observed slightly off this maximum, where the thermal conductivity is strongly suppressed.

For the thermoelectric figure of merit $ZT=S^2TG/(K^e+ K^{\rm ph})$ 
also the phononic contribution to the thermal conductivity $K^{\rm ph}$ 
has to be considered.
For a reduced electronic (thermal) conductivity, this phononic contribution 
is dominating. 
%Since the double quantum dot is not a bulk material, but a nanoscopic structure,
In contrast to bulk materials, the thermal conductivity of nanostructures %as quantum dots 
can be strongly suppressed. This is exploited in phonon engineering\cite{BiTePh}
which led to a historic breakthrough of higher figure of merit $ZT$
achieved in nanostructured materials, e.g.\ for quantum dot lattices\cite{Herman02}. On general grounds, one can expect a suppressed phononic contribution
to the  thermal  conductivity for nanostructures below the typical phonon mean free path. For semiconductors such as Si and GaAs this  mean free path is 0.1-1 $\mu$m,\cite{MFP} such that %one can expect a suppression of 
the phononic thermal conductivity is expected to be small.

In the case of the  molecule  instead of the quantum dot realization, 
the phonon contribution has been estimated in
Ref.\ \onlinecite{CK} to be between  $K^{\rm ph}=0$ (best case, no phonons)
and   $K^{\rm ph}=\pi^2 k_B T$ (worst case, maximal phonon contribution),
resulting in  $ZT\gtrsim 1$ and  $ZT\sim 2 \times 10^{-2}$, respectively, for
the attractive (negative) $U$ single quantum dot.
Due to the aforementioned particle-role relation, similar values for an 
optimal figure of merit are to be expected for our case of  a double dot (or molecule) with repulsive  $U$;  the actual $ZT$ of the presented calculations which are based
on the experimentally realized quantum dots\cite{Huebel1,Huebel2} and  $T=25\,$mK, instead of
parameters optimizing the  thermoelectric properties, are smaller.
Let us also remark that comparable efficiencies can be obtained  also
for non-interacting quantum dots, albeit at temperatures $T\sim {\cal O}(\Gamma_L+\Gamma_R)$, while for the Kondo quantum dots the maximal $ZT$ is at  $T\sim {\cal O}(T_K)\ll \Gamma_L+\Gamma_R$.

\section{Conclusion and outlook} We have shown that a
double quantum dot in the Kondo regime can be employed as 
a thermoelectric power generator. Our calculations are based
on the experiment of Ref.\ \onlinecite{Huebel1,Huebel2} whose focus was %to demonstrate the Kondo effect
the demonstration of Kondo correlations in the electric transport, not to optimize the thermoelectric properties.
Hence, %there is still
the study of different parameter regimes offers 
plenty of room to improve the thermopower and thermoelectric figure of merit. In fact, the relevant energy scales of the double quantum dot considered were in the $\mu$eV and mK regime, which makes these particular quantum dots unpractical for applications, except for Peltier cooling at ultra low temperatures in the mK
regime which might be of interest for basic research devices.
However, reducing the size of the quantum dots from $\mu$m towards nm,
the corresponding energy and temperature scales will be simply rescaled, as $ZT$. %A dramatic enhancement of the figure of merit is envisaged 
This can be pushed to even smaller sizes 
by employing molecules instead of quantum dots in molecular electronics\cite{Gimzewski87a}. 
In view of future applications, 
multiple quantum dots connected through a common back electrode (joint drain in Fig.\ \ref{Fig:Scheme}) provide a scalable setup, 
in which the generated power can be harvested through wires connecting the p- and n-type quantum dots separately. 

Possibly most promising is the integration of  the double quantum dot
thermoelectric element  on computer chips. %On one hand,  
Here the 
complex and otherwise expensive semiconductor device fabrication with
photolithographic and chemical processing is employed anyhow.
%On the other hand 
At the same time, the power consumption and cooling of waste heat
has become a critical issue for computer chip design. 
%For future designs  
On-chip cooling e.g.\ through liquid-filled microchannels is presently 
explored as a possible solution\cite{onchipcooling}, which would
involve additional technology and processing steps.
Here, the proposed double quantum dot device is much simpler.
\vspace{.5cm}

{\it Acknowledgments.}
We thank T. Costi, A.  H\"ubel, T. Pruschke, and J.\ Weis for discussions, the Austrian Science Fund (FWF) through SFB ViCom F41 
(SA, KH) and the European Research Council under the European Union's Seventh Framework Program (FP/2007-2013)/ERC 
through grant agreement n.\ 306447 (KH)  for financial support.

\end{document}